\documentclass{article}

\usepackage{arxiv}

\usepackage[utf8]{inputenc} 
\usepackage[T1]{fontenc}    
\usepackage{hyperref}       
\usepackage{url}            
\usepackage{booktabs}       
\usepackage{amsfonts}       
\usepackage{nicefrac}       
\usepackage{microtype}      
\usepackage{lipsum}
\usepackage{graphicx}

\usepackage{amsmath}
\usepackage{graphicx}
\usepackage{psfrag,epsf}
\usepackage{enumerate}
\usepackage{natbib}
\usepackage{amsfonts}
\usepackage{multirow}
\usepackage{float}
\usepackage{pifont}
\newcommand{\cmark}{\ding{51}}%
\newcommand{\xmark}{\ding{55}}%
\usepackage{url} 
\usepackage{enumerate}
\newtheorem{theorem}{Theorem}

\newcommand{\mbf}[1]{\mathbf{#1}}
\newcommand{\bx}{\mbf{x}}
\newcommand{\by}{\mbf{y}}

\newcommand{\bb}{\mbf{b}}
\newcommand{\bB}{\mbf{B}}
\newcommand{\bz}{\mbf{z}}
\newcommand{\be}{\mbf{e}}

\newcommand{\tB}{\text{B}}
\newcommand\numberthis{\addtocounter{equation}{1}\tag{\theequation}}

\usepackage{xcolor}

\title{A Transformation-free Linear Regression for Compositional Outcomes and Predictors}

\author{
 Jacob Fiksel \\
  Department of Biostatistics\\
  Johns Hopkins University \\
  Baltimore,
 MD 21205, USA \\
  \texttt{jfiksel@gmail.com} \\
   \And
 Scott Zeger \\
  Department of Biostatistics\\
  Johns Hopkins University \\
  Baltimore,
 MD 21205, USA \\
  \And
 Abhirup Datta \\
  Department of Biostatistics\\
  Johns Hopkins University \\
  Baltimore,
 MD 21205, USA \\
}

\begin{document}
\maketitle
\begin{abstract}
Compositional data are common in many fields, both as outcomes and predictor variables. The inventory of models for the case when both the outcome and predictor variables are compositional is limited and the existing models are difficult to interpret, due to their use of complex log-ratio transformations. We develop a transformation-free linear regression model where the expected value of the compositional outcome is expressed as a single Markov transition from the compositional predictor. Our approach is based on generalized method of moments thereby not requiring complete specification of data likelihood and is robust to different data generating mechanism. Our model is simple to interpret,  allows for 0s and 1s in both the compositional outcome and covariates, and subsumes several interesting subcases of interest. We also develop a permutation test for linear independence. Finally, we show that despite its simplicity, our model accurately captures the relationship between compositional data from education and medical research.
\end{abstract}


\section{Introduction}
\label{sec:intro}

Compositional data, also referred to as fractional data \citep{mullahy2015,murteira2016}, consist of vectors constrained to lie in the unit simplex, $\mathbb{S}^{D}$, where $\mathbb{S}^{D} = \{ (x_{1}, x_{2}, \ldots, x_{D})^{'}| x_j \geq 0, i=j,\ldots, D; \sum_{i=j}^{D}x_{j} = 1\}$. Compositional data appear in many fields, such as econometrics \citep{papke1996}, geochemistry \citep{templ2008}, physical activity research \citep{dumuid2018}, microbiome analysis \citep{lin2014}, and nutritional epidemiology \citep{leite2016}. 

Depending on the application, compositional data may appear as an explanatory variable \citep{hron2012, mcgregor2019, dumuid2018}, as an outcome of interest \citep{papke1996, mullahy2015, egozcue2012, hijazi2009}, or both \citep{wang2013, chen2017, alenazi2019}. While there has been much attention placed on the first two cases, little work has been done on creating simple and interpretable models for the last case. Examples of problems with both compositional outcomes and explanatory variables include relating the percentage of males and females with different education levels across countries \citep{appliedCDA}, modeling the relationship between age structure and consumption structure across economic areas \citep{chen2017}, and understanding how different methods for estimating the composition of white blood cell types are related \citep{aitchison1986, alenazi2019}.

All current methods developed specifically for problems where both the outcome and the explanatory variable are compostional require data transformation. 
\citet{chen2017} transforms both the response and explanatory compositional variables, while \citet{alenazi2019} transforms just the compositional explanatory variable. Transformation based models limit interpretability \citep{morais2018}, especially when complex, but commonly used transformations such as the isometric log-ratio (ILR) transformation \citep{egozcue2003} are used. Furthermore, many transformations do not allow for compositional data with 0s and 1s \citep{appliedCDA}.

In this manuscript, we postulate a simple estimating equation that directly relates the expected value of the compositional outcome as a linear function of the compositional explanatory variable. Our approach does not require any transformation of the data and naturally accommodates 0s and 1s, thus treating data on the interior of the simplex the same as data on the boundary. By linearly relating the outcome and explanatory variables, the parameters in our model are easily interpretable, unlike transformation based compositional regression models. We develop an expectation-maximization (EM) \citep{dempster1977} algorithm for fast and accurate parameter estimation via constrained maximization of the quasi-likelihood that respects the unit sum nature of the compositional data. We present simulation results comparing the models for compositional data under a variety of data generating mechanisms. We also present a permutation-based test for assessing whether or not there exists a linear dependency between the outcome and explanatory variables, and evaluate the operating characteristics of this test via simulation. Finally, we demonstrate the utility of our model with two data analyses from education and medical research.

\section{Review of Transformation Based Compositional Regression Models}
\label{sec:review}

Current models for problems with compositional outcomes and explanatory variables rely on transforming the compositional data from $\mathbb{S}^{D}$ to $\mathbb{R}^{D-1}$. The recommended transformation for compositional data is the ILR transformation \citep{egozcue2003, hron2012, appliedCDA}, where for $\bz \in \mathbb{S}^{D}$

\begin{equation*}\label{eq:ilr}
    ilr(\mathbf{z})_{j} = \sqrt{\frac{D-j}{D-j+1}}ln\left(\frac{z_{j}}{\left(\prod_{k=j+1}^{D}z_{k}\right)^{\frac{1}{D-j}}} \right),\ j=1,\ldots, D-1.
\end{equation*}
The mathematical advantage of using the ILR transformation over more simple transformations, such as the additive log-ratio (ALR) or centered log-ratio (CLR) \citep{aitchison1986}, is that the vector $ilr(\bz)$ can be used as covariates in a standard linear regression model without having to constrain the regression coefficients \citep{hron2012}.

The model presented by \citet{chen2017} assumes that for an outcome $\by \in \mathbb{S}^{D_{r}}$ and explanatory variable $\bx \in \mathbb{S}^{D_{s}}$, where $D_{r}$ is not necessarily equal to $D_{s}$, that

\begin{equation}\label{eq:ilrreg}
    E[ilr(\by)_{k} | \bx] = \beta_{0k} + \sum_{j=1}^{D_{s}-1}\beta_{jk}ilr(\bx)_{j},\ k=1,\ldots, D_{r}-1.
\end{equation}
Hence, $\beta_{11}$ has an interpretation as the effect of increasing the relative value of $x_{1}$ by 1 compared to the rest of $\bx$, holding the ratios between the other components of $\bx$ constant, on the change of the relative value of $y_{1}$ compared to the rest of $\by$; the other regression coefficients have no meaningful interpretation \citep{hron2012, chen2017}. To obtain the effects of relative changes of each part of $\bx$ on $\by$, one must use the permutation operation,
\begin{equation*}\label{eq:pivot}
    \bz^{l} = (z_{l}, z_{1}, \ldots, z_{l-1}, z_{l+1},\ldots, z_{D}),
\end{equation*}
and estimate $D_{r} \cdot D_{s}$ separate models where
\begin{equation}\label{eq:ilrregpivot}
\resizebox{0.91\hsize}{!}{%
   $ E[ilr(\by^{l_{1}})_{k}] = \beta_{0k}^{(l_{1}, l_{2})} + \sum_{j=1}^{D_{s}-1}\beta_{jk}^{(l_{1}, l_{2})}ilr(\bx^{l_{2}})_{j},\ k=1,\ldots, D_{r}-1,\ l_{1}=1,\ldots, D_{r},\ l_{2}=1,\ldots, D_{s}$%
    }.
\end{equation}

The coefficients of interest would then be $\beta_{11}^{(l_{1}, l_{2})}$ for each combination of $l_{1}$ and $l_{2}$ \citep{chen2017, appliedCDA}. As parameter estimation is performed using standard maximum likelihood for linear regression models, this procedure is not  computationally expensive. However, using multiple versions of a model to obtain a set of coefficients that cannot be interpreted jointly is undesirable. 
There are two additional downsides. First, the ILR transformation does not allow for 0s in the compositional data. If either $\bx$ or $\by$ are categorical, the ILR transformation framework can not be used, even though categorical variables are still in the unit simplex. Second, the coefficients of interest can only be vaguely interpreted in terms of changes in the relative values of each part of the compositional data to the geometric mean. This model does not permit for simple interpretation of the coefficients in terms of the direct effect of changing the value of $\bx$ within the simplex on the expected value of $\by$ in the simplex \citep{morais2018}. The lack of a simple interpretation for the coefficients in (\ref{eq:ilrregpivot}) have forced practitioners to instead rely on graphical techniques to display the estimated response surface of $\by$ as a function of $\bx$ \citep{nguyen2018}. 

\citet{alenazi2019} takes a different approach to compositional regression, as only the explanatory compositional variable $\bx$ is transformed. While \citet{alenazi2019} is more interested in prediction accuracy than interpretation and uses a complex principal components based transformation, one can use any transformation $t$ (e.g., the ILR transformation). The assumed regression model is the multinomial logit specification \citep{papke1996, mullahy2015, murteira2016}:

\begin{align}\label{eq:alenazi}
\begin{split}
E[y_{k} | \bx] &= \frac{\text{exp}(\beta_{0k} + \sum_{j=1}^{D_{s}-1}\beta_{jk}t(\bx)_{j})}{1 + \sum_{k=1}^{D-1}\left[\text{exp}(\beta_{0k} + \sum_{j=1}^{D_{s}-1}\beta_{jk}t(\bx)_{j})\right]},\ k=1,\ldots, D_{r}-1\\
E[y_{D_{r}} | \bx] &= \frac{1}{1 + \sum_{k=1}^{D-1}\left[\text{exp}(\beta_{0k} + \sum_{j=1}^{D_{s}-1}\beta_{jk}t(\bx)_{j})\right]}.
\end{split}
\end{align}

\citet{murteira2016} discuss both quasi-maximum and maximum likelihood (QML and ML) methods for estimation of the coefficients. However, \citet{alenazi2019} uses a QML method which allows for 0 values in $\by$ \citep{papke1996, mullahy2015, murteira2016}, and does not make any distributional assumptions about $\by$. 




 Despite this method allowing for potential 0s in $\by$ (and in $\bx$ if one uses a transformation that allows for 0s, such as the $\alpha$-transformation \citep{tsagris2015regression}), the regression coefficients are still only interpretable in terms of effects of changing a transformed version of $\bx$ on $\log\left(\frac{E[y_{j}]}{E[y_{D_{r}}]} \right)$. In order to interpret the model in terms of changes within the simplex, one would again need to resort to graphical techniques.

\section{Direct Regression of Compositional Variables on the Simplex}\label{sec:compreg}

Section \ref{sec:review} showed that current models for regressing a compositional outcome on a compositional explanatory variable are difficult to interpret due to modeling transformed versions of the compositional data. To create an interpretable model for this class of problems, we want to directly model the expected value of $\by$ as a linear function of $\bx$. This is achieved through the following linear model:

\begin{equation}\label{eq:directreg}
    E[\by | \bx] = \sum_{j=1}^{D_{s}} x_{j}\bb_{j},
\end{equation}
where $\bb_j$'s are $D_y$-dimensional vectors. Letting $\bB$ represent the matrix with the $jth$ row $\bB_{j*} = \bb_{j}^{'}$, we can rewrite the model in (\ref{eq:directreg}) as
\begin{equation}\label{eq:markov}
    E[\by | \bx] = \bB^{'}\bx\ .
\end{equation}

Because $\by$ is compositional, we require that $\sum_{k=1}^{D_{r}}E[y_{k} | \bx] = 1$. To adhere to the unit sum restriction, we take advantage of the fact that $\bx$ is also compositional. Hence, it suffices to constrain $\bB$ to be a Markov (transition) matrix with non-negative entries and rows summing to 1, i.e., 

\begin{equation*}\label{eq:constraints}
    \bB \in \{\mathbb{R}^{D_{s} \times D_{r}} | B_{jk} \geq 0, \sum_{k=1}^{D_{r}}\text{B}_{jk} = 1\ \text{for } j=1,\ldots, D_{s} \} \ .
\end{equation*}

This transformation-free model allows 0s and 1s in both $\bx$ and $\by$ as (\ref{eq:markov}) is well-defined for entire $\bx$- and $\by$-simplexes including the boundaries. The model  allows for direct interpretation of the association between $\bx$ and $E[\by]$ in terms of the regression coefficient matrix $\bB$. If $x_{j}$ increases by $\Delta \in (0, 1-x_{j}]$, at the expense of $x_{k}$ decreasing by $\Delta$ (assuming $x_{k} \geq \Delta$) and holding the rest of $\bx$ constant, the expected change in $E[\by]$ is expressed as $\Delta(\bB_{j*} - \bB_{k*})$. This interpretation respects the fact that increasing one part of $\bx$ necessarily involves the trade-off of decreasing at least one other part of $\bx$. For example, if $\bx$ represents the proportion of each day spent on different activities such as sleep, physical activity, and sedentary time, we may be interested in how components of a compositional $\by$ are expected to change when we increase physical activity and decrease sedentary time. We also may be interested in how this compares to the change of $\by$ when we instead increase physical activity at the expense of sleep \citep{dumuid2018}. Another example application where this interpretation is useful is in marketing, where teams may want to know whether to increase the percentage of expenditure on television advertisements at the expense of radio advertisements or press advertisements in order to best increase their market share \citep{morais2018}. Furthermore, our model allows us to directly estimate how the expected value of $\by$, rather than some transformed version of $\by$, is associated with changes in $\bx$.

In addition to the simple interpretation, the direct regression model in (\ref{eq:directreg}) exhibits other convenient statistical properties. First, consider the case when two rows, $j_{1}$ and $j_{2}$, of $\bB$ are equal. This implies that increasing $x_{j_1}$ at the expense of $x_{j_2}$ does not change $E[\by]$. We then have
\begin{align*}
    E[\by | \bx] &= \sum_{j \neq j_{1}, j_{2}}^{D_{s}}x_{j}\bb_{j} + x_{j_1}\bb_{j_1} + x_{j_2}\bb_{j_2}\\
    &= \sum_{j \neq j_{1}, j_{2}}^{D_{s}}x_{j}\bb_{j} + \bb_{j_1}(x_{j_1} + x_{j_2})\ \numberthis \label{eq:equalrows},
\end{align*}
which shows that we can treat the combined categories $x_{j_1} + x_{j_2}$ as a single category. This not only simplifies interpretation of the direct regression model, but also means that there is one less row of $\bB$ to estimate. 

Similarly, the direct regression model can easily accommodate combining categories $y_{k_1}$ and $y_{k_2}$. The direct regression model implies that
\begin{align*}
    E[y_{k_1} + y_{k_2} | \bx] &= \sum_{j=1}^{D_{s}}B_{jk_{1}}x_{j}  + \sum_{j=1}^{D_{s}}B_{jk_{2}}x_{j} \\
    &=\sum_{j=1}^{D_{s}}(B_{jk_{1}}+ B_{jk_{2}})x_{j}.
\end{align*}
Thus, conditional expectations of linear combinations of $\by$ can be obtained through adding columns of $\bB$. This ensures that the model is invariant to aggregating outcome categories. Rather than having to perform separate regressions for different choices of aggregation of the outcome categories, practitioners can simply perform one regression using the full set of categories, and aggregate columns of $\bB$ post-hoc.

Because $\bB$ is a Markov matrix, the rows of $\bB$ are themselves members of $\mathbb{S}^{D_{r}}$. If we let $x_{j}=1$, which means that $\bx$ is in the $j$th corner of $\mathbb{S}^{D_{s}}$, (\ref{eq:directreg}) shows that $E[\by | x_{j} = 1] = \bb_{j}$. Thus, $\bB_{j*}$ is equivalent to $E[\by]$ when $x_{j}=1$. For the case when $D_{r}=3$, this means we can actually visualize the coefficients themselves using a ternary diagram \citep{ggtern}. 
Consider the following two values of $\bB$: 

\begin{equation*}
    \bB^{(1)} = \begin{pmatrix}
.90 & .05 & .05\\
.05 & .90 & .05\\
.05 & .05 & .90
\end{pmatrix};\ \bB^{(2)} = \begin{pmatrix}
.40 & .30 & .30\\
.30 & .40 & .30\\
.30 & .30 & .40
\end{pmatrix}
\end{equation*}

\begin{figure}[H]
\centering
\includegraphics[width=\linewidth]{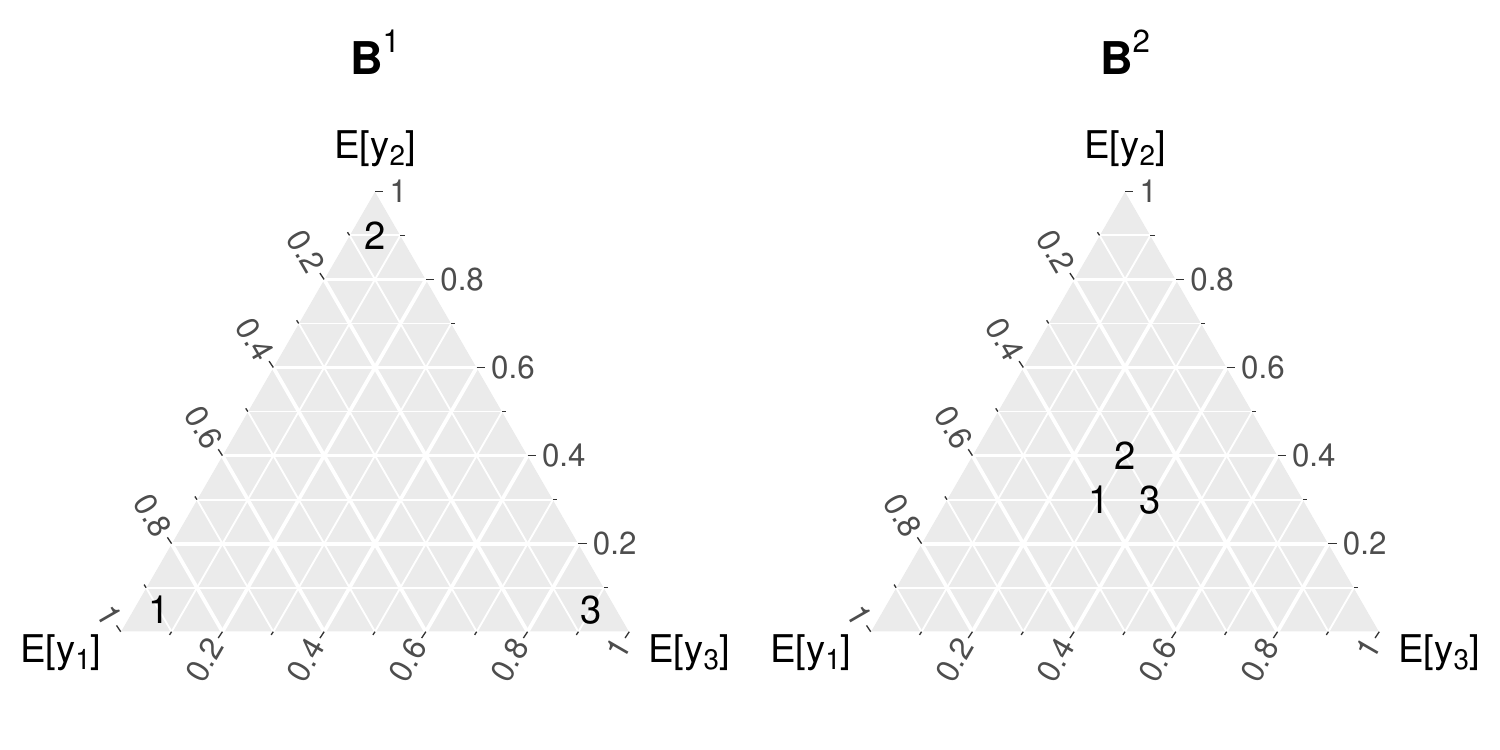}
\caption{Visualization of the coefficients $\bB$. For a number $j$, the point plots $\bB_{j*}$ within a ternary diagram.}\label{fig:bviz}
\end{figure}

$\bB^{(1)}$ represents the setting when $\by$ and $\bx$ are highly correlated, while $\bB^{(2)}$ represents  the setting when $\by$ and $\bx$ are weakly correlated. This interpretation is derived directly from the simple analytic interpretation of the direct regression model in (\ref{eq:directreg}). This interpretation is also seen through plotting the rows of these two matrices in a ternary diagram, as in Figure \ref{fig:bviz}. Each number in the plot corresponds to a row in the two values of $\bB$. The plot of $\bB^{(1)}$ shows that $E[\by]$ substantially changes with $\bx$, as changes in $E[\by]$ with $\bx$ can be expressed as scaled differences in the rows of $\bB$. However, the plot of $\bB^{(2)}$ shows much smaller changes for $E[\by]$ with $\bx$. Confidence regions for each row of $\bB$ can also be plotted within the diagram. We demonstrate this in the example in Section \ref{sec:educ}.

We note that the models of 
\citet{chen2017} and \citet{alenazi2019} models have some advantages over our simple and direct model, most notably the ability to include multiple confounding covariates of mixed variable type in the model, and we present a full comparison of the properties of each model in Table 1. However, the simple interpretation of the direct regression model stands in stark contrast to the vague interpretation of the coefficients in the ILR model or any model which transforms $\by$ and/or $\bx$. The interpretation of $\bB$ is simple to communicate to non-statisticians without graphical techniques, does not require familiarity with the compositional transformations, and only requires estimating one single model for $E[\by|\bx]$, rather than $D_{r} \times D_{s}$ models. The direct regression model also seamlessly permits 0s and 1s in both $\bx$ and $\by$, leading to the sub-cases of interest presented in Sections \ref{sec:catcov} and \ref{sec:catoutcome}. 

\begin{table}[t!]
\caption{Comparison of properties between the three compositional regression models. A \cmark \ indicates that a model has the given property, while a \xmark \ indicates that a model does not have the given property.}
\label{tbl:propertiescompare}
\centering
    \begin{tabular}{|p{0.5\textwidth}|p{0.15\textwidth}||p{0.15\textwidth}||p{0.15\textwidth}|}
    \hline
     Properties & Direct Regression & ILR transformation \citep{chen2017} & Multinomial logit  \citep{alenazi2019} \\ \hline
    Transformation-free & \cmark & \xmark & \xmark \\ \hline
    Accommodates 0s and 1s in both outcome and predictor compositions & \cmark & \xmark & \cmark \\ \hline
    Coefficients interpreted in terms of changes of $\by$ in the simplex & \cmark & \xmark & \xmark \\ \hline
    Only requires running 1 model, instead of $D_{r} \times D_{s}$ models & \cmark & \xmark & \cmark \\ \hline
    Coefficients interpreted in terms of changes of log ratios of $\by$& \xmark & \cmark & \cmark \\ \hline
    Can be extended to include multiple covariates that may be compositional, continuous, or discrete & \xmark & \cmark & \cmark \\ \hline
    \end{tabular}
\end{table}

\subsection{Categorical covariates}\label{sec:catcov}

For each observation $i$, assume that the covariate of interest is whether or not the observation belongs to one of $j=1,\ldots, D_{s}$ groups. If observation $i$ belongs to subgroup $j$, we let $\bx_{i} = \be_{j}$, where $\be_{j}$ is the compositional vector with a 1 in the $jth$ index. 
We now have an ANOVA-like model, but with a compositional outcome.

This model has been considered in the literature where only the outcome is compositional, but previous solutions have either used an ILR transformation for $\by$ \citep{appliedCDA} or assumed that $\by | \bx$ follows a Dirichlet distribution \citep{maier2014}. Our model allows for a transformation-free and distribution-free solution for this problem. The formulation of our model in (\ref{eq:directreg}) shows that $\bB_{j*} = E[\by | \bx = \be_{j}]$, i.e., the rows of $\bB$ simply interprets as the expectation for the $j^{th}$ group. If we are interested in how $E[\by]$ changes between two groups $j_1$ and $j_2$, this change is represented by $\bB_{j_1*} - \bB_{j_2*}$. 
If the rows of $\bB$ are all equal, this would indicate linear independence  between $\by$ and $\bx$.

\subsection{Categorical outcome}\label{sec:catoutcome}

We now $\by$ restrict to be categorical, meaning that each observation $i$ belongs to one of $k=1,\ldots, D_{r}$ groups. The standard model for this case would be a multinomial logistic model, using the ILR transformed $\bx$ as covariates \citep{appliedCDA}. However, we can use the model in (\ref{eq:directreg}), which allows for direct estimation of $E[y_{k} | \bx] = P(\by = \be_{k} | \bx),\ k=1,\ldots, D_{r}$. This is equivalent to performing multinomial linear regression, with an identity link. The identity link is the canonical link here, as the covariates are compositional. Further restricting $\bx$ to be categorical reduces this to a $D_r \times D_S$ contingency table. $\bB_{j_1, k_1}$ can be interpreted now as the conditional probability $P(\by = \be_{k_1} | \bx = \be_{j_1})$ and $\bB_{j_1, k_1} - \bB_{j_2,k_1}$ 
is the 
risk difference between groups.

\subsection{Discrete time series transition probabilities}\label{sec:ar1}

A specific case of a categorical outcome and covariate is in estimating time-invariant transition probabilities for a first-order Markov process. An example of this class of problems is estimating the probability of firms or institutions transitioning between specific credit ratings \citep{jones2005}. Observations may transition between $r=1,\ldots, R$ states. In the ideal case, for each observation unit $i$, we observe their discrete state $\by_{i,t}$ over times $t=0,\ldots, T$. We are then interested in estimating the probability that each observation moves to state $j$ at time $t$, given that they are in state $k$ at time $t-1$ (assuming transition probabilities are constant over time and between observation units). The interpretation of $\bB$ from Sections \ref{sec:catoutcome} and \ref{sec:catcov} shows that if the covariate in (\ref{eq:directreg}), $\by_{i,t-1}$, and the outcome is, $\by_{i,t}$, 
then $\bB_{jk} = P(\by_{i,t} = \be_{j} | \by_{i,t-1} = \be_{k})$, which is exactly the transition probability we seek to estimate. The estimation procedure we outline in Section \ref{sec:estimation} will then coincide with the MLE of $\bB$. 

\subsection{AR(1) model for compositional data}
Rather than observing the states of each observation unit, we may only observe the percentage of observations in each state at each time. For example, \citet{jones2005} presents the case where for each year between 1984-2004, we only observe the percentage of commercial banks that belong to four different categories of credit quality. Our observed data is now the percentage of units in the different states at time $t$, $\by_{t}$. Specifically, $y_{tj}$ is the percentage of observations belonging to state $j$ at time $t$. \citet{lee1970}, \citet{macrae1977}, and \citet{jones2005} have shown that $E[\by_{t} | \by_{t-1}] = \bB^{'}\by_{t-1}$, where $\bB_{ij}$ is again defined as $P(\by_{i,t} = \be_{j} | \by_{i,t-1} = \be_{k})$. Thus, the direct regression model in (\ref{eq:markov}) can be used to estimate the individual transition probabilities, despite only observing aggregate data. For such settings, our model can be perceived as an AR(1) model for the compositional time series $y_t$. 

\section{Parameter Estimation}\label{sec:estimation}

\subsection{Generalized Method of Moments Approach}

In order to estimate the entries of $\bB$, we note that the model in (\ref{eq:markov}) implies that 
\begin{equation*}\label{eq:marginalmean}
    E[y_{k} | \bx] = \sum_{j=1}^{D_{s}}B_{jk}x_{j} \ .
\end{equation*}
As we are only interested in the first moment of $\by | \bx$, we use a generalized method of moments (GMM) \citep{hansen1982} approach and seek a function $\ell(\bB; \by, \bx)$ such that
\begin{equation*}
      E_{\bB_0} \left(\frac {d \ell}{d \bB} \right) = 0\ ,
\end{equation*}
where $\bB_{0}$ is the true value of $\bB$. A function $\ell$ which achieves this, while also allowing for 0s in $\by_{i}$ and $\bx_{i}$, is the Kullback-Leibler distance (KLD) between two compositional vectors --- the observed $\by_{i}$ and $E[\by_{i} | \bx_{i}]$ \citep{GBQL} --- i.e., 

\begin{align*}\label{eq:compregkld}
\ell&= \sum_{i=1}^{N} \mbox{KLD}(y_i \;\|\; E[y_i \;|\; x_i]) \\
&= -\sum_{i=1}^{N}\sum_{k=1}^{D_{r}}y_{ik}\log\left(\frac{E[y_{ik}| \bx]}{y_{ik}}\right)\\
   &= -\sum_{i=1}^{N}\sum_{k=1}^{D_{r}}y_{ik}\log\left(\frac{\sum_{j=1}^{D_{s}}B_{jk}x_{ij}}{y_{ik}}\right)\ . \numberthis
\end{align*}

 Letting $\mathcal{F} = \{ \bB;B_{jk} \geq 0, \sum_{k=1}^{D_{r}}\text{B}_{jk} = 1\}$ be the constrained space for $\bB$, minimizing (\ref{eq:compregkld}) with respect to $\bB$ is equivalent to maximizing the log-quasi-multinomial likelihood \citep{mullahy2015, alenazi2019}:

\begin{align*}\label{eq:objfn}
 \min_{\bB \in \mathcal{F}}\ell(\bB; \bx, \by) &=
    \min_{\bB \in \mathcal{F}} -\sum_{i=1}^{N}\sum_{k=1}^{D_{r}}y_{ik}\log\left(\frac{\sum_{j=1}^{D_{s}}B_{jk}x_{ij}}{y_{ik}}\right)\\ &=
    \max_{\bB \in \mathcal{F}}\sum_{i=1}^{N}\sum_{k=1}^{D_{r}}y_{ik}\log\left(\sum_{j=1}^{D_{s}}B_{jk}x_{ij}\right) \numberthis
\end{align*}

 The multinomial quasi-likelihood belongs to the linear exponential family \citep{gourieroux1984} and minimizing (\ref{eq:compregkld}) (or equivalently, maximizing (\ref{eq:objfn})) produces a consistent estimator for $\bB_0$ \citep{gourieroux1984, papke1996, mullahy2015}. When $\by$ is categorical (examples in Sections \ref{sec:catoutcome} and \ref{sec:ar1}), the quasi-likelihood becomes the proper likelihood for multinomial distribution and the estimate of $\bB$ becomes the MLE. More generally for compositional $\by$ and $\bx$, \citet{GBQL} show that (\ref{eq:compregkld}) is convex with respect to $\bB$, guaranteeing existence of a global minimum of (\ref{eq:compregkld}).




\subsection{An EM Algorithm for Maximizing the Objective Function}

\citet{alenazi2019} also uses a GMM approach via minimization of the KLD between the observed and expected values for the compositional outcome in (\ref{eq:alenazi}). Because the form of the conditional expected value in (\ref{eq:alenazi}) is that used in multinomial logistic regression, the coefficients are unconstrained and  \citet{alenazi2019} utilizes the Newton-Raphson \citep{bohning1992multinomial} algorithm for maximizing the log-quasi-multinomial likelihood. However, our model imposes constraints on the parameter space for $\bB$ making it difficult to employ the Newton-Raphson algorithm to maximize (\ref{eq:objfn}). 

We instead develop an EM algorithm for parameter estimation by maximization of (\ref{eq:objfn}). We first present the algorithm for the special case where $\by_{i}$'s are categorical (Section \ref{sec:catoutcome}). We introduce ``missing'' pseudo categories $\bx_{i}^{*}$ such that $\bx_{i}^{*} | \bx_{i} \sim Multinomial(1, \bx_{i})$ and assume $\by_{i} | \bB, x_{ij}^{*} = 1 \sim Multinomial(1, \bB_{j*})$, thus using a proper likelihood for the outcome. We then arrive
at the following likelihood of $\by | \bx$ (marginalizing out the psuedo-categories $\bx^{*}$):

\begin{align*}
    p(\by | \bB, \bx) &= \prod_{i=1}^{N}\left(\sum_{j=1}^{D_{s}}p(x_{ij}^{*}=1)p(\by_{i}^{*} |\bB, x_{ij}^{*} = 1)\right)\\
    &= \prod_{i=1}^{N}\left(\sum_{j=1}^{D_{s}}x_{ij}\prod_{k=1}^{D_{r}}(B_{jk})^{y_{ik}}\right)\\
    &= \prod_{i=1}^{N}\prod_{k=1}^{D_{r}}\left(\sum_{j=1}^{D_{s}}B_{jk}x_{ij}\right)^{y_{ik}}\numberthis\label{eq:pseudomarginal}
\end{align*}

Taking the log of (\ref{eq:pseudomarginal}) gives us the form of the objective function in (\ref{eq:objfn}). Letting $\text{B}_{jk}^{(t)}$ denote the value of $\tB_{jk}$ after iteration $t$, the expected complete log-likelihood becomes

\begin{equation*}
    Q(\bB | \bB^{(t)}) = \sum_{i=1}^{N}\sum_{j=1}^{D_2}\left[E[x_{ij}^{*} | x_{ij}, y_{ik}, \tB_{jk}^{(t)}](log(x_{ij}) + \sum_{k=1}^{D_1} y_{ik}log(\tB_{jk}))\right].
\end{equation*}

Noting that the M-step will require finding 

\begin{equation}\label{eq:mstepinit}
      \max_{\bB \in \mathcal{F}}\sum_{i=1}^{N}\sum_{k=1}^{D_{r}}\sum_{j=1}^{D_{s}}E[x_{ij}^{*} | x_{ij}, y_{ik}, \tB_{jk}^{(t)}]y_{ik}\log(\tB_{jk}) \ ,
\end{equation}
we see that the terms in (\ref{eq:mstepinit}) for which $y_{ik}=0$ will not influence the maximization. Thus, rather than evaluating both $E[x_{ij}^{*} | x_{ij}, y_{ik} = 0, \tB_{jk}^{(t)}]$ and $E[x_{ij}^{*} | x_{ij}, y_{ik} = 1, \tB_{jk}^{(t)}]$, we only have to evaluate the latter term. We thus introduce weights $\pi_{ijk}^{(t+1)}$ for the E-step at iteration $t+1$ which are equal to $E[x_{ij}^{*}| x_{ij}, y_{ik} = 1, \tB_{jk}^{(t)}]$:

\begin{equation*}\label{eq:estep}
    \pi_{ijk}^{(t+1)} = \frac{x_{ij}\tB_{jk}^{(t)}}{\sum_{j=1}^{D_{s}}x_{ij}\tB_{jk}^{(t)}},\ i=1,\ldots, N, j=1,\ldots, D_{s}, k=1,\ldots, D_{r} \ .
\end{equation*}

The expected complete log-likelihood is now
\begin{equation*}\label{eq:newqstep}
    Q(\bB | \bB^{(t)}) =\sum_{i=1}^{N}\sum_{k=1}^{D_{r}}\sum_{j=1}^{D_{s}}y_{ik}\pi_{ijk}^{(t+1)}\log(\tB_{jk})\ ,
\end{equation*}
and the M-step from (\ref{eq:mstepinit}) becomes 
\begin{equation}\label{eq:mstepobj}
\max_{\bB \in \mathcal{F}}Q(\bB | \bB^{(t)}) = \max_{\bB \in \mathcal{F}}\sum_{i=1}^{N}\sum_{k=1}^{D_{r}}\sum_{j=1}^{D_{s}}y_{ik}\pi_{ijk}^{(t+1)}\log(\tB_{jk})\ .
\end{equation}

Due to the fact that $\sum_{k=1}^{D_{r}}\tB_{jk}=1$ for $j=1,\ldots, D_{s}$, we can recognize the constrained maximization in (\ref{eq:mstepobj}) equivalent to maximizing $j=1,\ldots, D_{s}$ weighted multinomial likelihoods. This implies the following M-step:

\begin{equation*}\label{eq:mstepB}
    \tB_{jk}^{(t+1)} = \frac{\sum_{i=1}^{N}y_{ik}\pi_{ijk}^{(t+1)}}{\sum_{k=1}^{D_{r}}\sum_{i=1}^{N}y_{ik}\pi_{ijk}^{(t+1)}},\ k=1,\ldots, D_{r},\ j=1,\ldots, D_{s}\ .
\end{equation*}

Having developed an EM algorithm when we restrict the outcome $\by$ to be categorical, Theorem 1 now extends the EM algorithm to the general case when $\by$ is compositional:

\begin{theorem}
Let $f(t) = \sum_{i=1}^{N}\sum_{k=1}^{D_{r}}y_{ik}\log\left(\sum_{j=1}^{D_{s}}B_{jk}^{(t)}x_{ij}\right)$ be the value of the objective function after iteration $t$ of the EM algorithm with compositional outcomes $\by$, using the same E and M steps as when $\by$ is categorical. Then $f(t+1) - f(t) \geq 0$, with strict inequality if $Q(\bB^{(t+1)}| \bB^{(t)}) > Q(\bB^{(t)}| \bB^{(t)})$.
\end{theorem}

A proof is provided in Web Appendix A. Theorem 1 allows use of the same EM algorithm for estimation of $\bB$, despite the fact that our approach is likelihood-free and only specifies $E[\by | \bx]$. As both the E-step and M-steps are available in closed form, the implementation of this EM-algorithm is extremely fast. The EM-algorithm can be further accelerated through use of the SQUAREM R-package \citep{squarem}.

\section{A permutation test for linear independence}\label{sec:indeptest}

In the \citet{chen2017} and \cite{alenazi2019} models presented in (\ref{eq:ilrreg}) and (\ref{eq:alenazi}), one can test whether each of the coefficients is equal to 0, using either bootstrapping \citep{efron1994} or by estimating the standard errors of the coefficient estimates \citep{chen2017, mullahy2015}. This is testing whether certain parts of $\by$ and $\bx$ are associated with each other. We now present a permutation test for linear independence that can be applied to the direct regression method, and also can be adapted to the \citet{chen2017} and \cite{alenazi2019} models. 

If $\by$ is linearly independent of $\bx$, we have $E[\by | \bx] = E[\by]$. The interpretation of our model in Section \ref{sec:compreg} shows that this is equivalent to restricting the model in (\ref{eq:markov}) such that the rows of $\bB$ are equal. We now develop a procedure for testing the following null hypothesis:

\begin{equation*}\label{eq:h0}
    H_{0}: E[\by] = \bB_{1*}=\bB_{2*}=\cdots=\bB_{D_{r}*} \ .
\end{equation*}

Letting $\boldsymbol{\mu} = E[\by]$, under the restricted model implied by $H_{0}$, the maximization task in (\ref{eq:objfn}) becomes

\begin{equation}\label{eq:indepobjfn}
    \max_{\boldsymbol{\mu} \in \mathbb{S}^{D_{r}}}\sum_{i=1}^{N}\sum_{k=1}^{D_{r}}y_{ik}\log(\mu_{k})\ .
\end{equation}.
The solution to the constrained maximization task in (\ref{eq:indepobjfn}) leads to the following estimate of $\boldsymbol{\mu}$:
\begin{equation*}\label{eq:indepest}
    \hat{\boldsymbol{\mu}} = \frac{1}{N}\sum_{i=1}^{N}\by_{i}\ ,
\end{equation*}
which is simply the arithmetic average of the observed $\by$. Letting $\bar{y}_{k} = \frac{1}{N}\sum_{i=1}^{N}y_{ik}$, under $H_{0}$ the log-quasi likelihood in (\ref{eq:objfn}) becomes
\begin{equation*}\label{eq:indeppll}
    PLL_{H_{0}}=\sum_{i=1}^{N}\sum_{k=1}^{D_{r}}y_{ik}\log(\bar{y}_{k})\ .
\end{equation*}

Under the alternative hypothesis,
\begin{equation*}\label{eq:h_alt}
    H_{A}: \bB_{1*} \neq \bB_{k*} \text{ for at least one value of } k \in \{2,\ldots, D_{r}\} \ ,
\end{equation*}
the log-quasi likelihood is that implied in (\ref{eq:objfn}):
\begin{equation*}\label{eq:fullpll}
    PLL_{H_{A}}=\sum_{i=1}^{N}\sum_{k=1}^{D_{r}}y_{ik}\log\left(\sum_{j=1}^{D_{s}}\hat{B}_{jk}x_{ij}\right) \ .
\end{equation*}

Comparing the log-quasi likelihoods under $H_{0}$ and $H_{A}$ leads to the following test statistic of interest:

\begin{equation*}\label{eq:llrstat}
    \lambda = PPL_{H_{A}} - PPL_{H_{0}}
\end{equation*}
which is equivalent to the log-quasi likelihood ratio between the restricted and full models. To obtain the distribution of $\lambda$ under $H_{0}$, we use the following Monte Carlo permutation testing procedure \citep{good2005}:

\begin{enumerate}[Step 1:]
\item Obtain $\lambda^{obs}$ using the observed $\bx$ and $\by$. 
\item Randomly permute the entries of $\bx$ as under $H_0$ $y$ and $x$ are linearly independent 
\item Obtain $\lambda^{perm}$ using the permuted $\bx$ and observed $\by$.
\item Repeat Steps 2-3 $b=1,\ldots,B$ times,  obtaining $\lambda^{perm_{b}}$ for each permutation. In practice, setting $B=1000$ appears to give good precision \citep{zeng2015}. 
\item Calculate the p-value, $p=\frac{1}{B}\sum_{b=1}^{B}I(\lambda^{perm_{b}} \geq \lambda^{obs})$
\end{enumerate}

The permutation test procedure allows for testing of whether changing any part of the compositional $\bx$ is associated with a linear change in the expected value of $\by$.  Furthermore, this procedure can be adopted for use in the models presented by \citet{chen2017} and \citet{alenazi2019}, either using the normal likelihood for the ILR transformed outcome, or the log-quasi likelihood using the conditional expected value formulation in (\ref{eq:alenazi}). 

\section{Simulation studies}

\subsection{Model comparison study}\label{sec:modelcompare}

We first perform simulations to compare the performance of the direct regession model with that of the \citet{chen2017} model and the \citet{alenazi2019} model across situations when only one of the three models is correctly specified. To generate realistic data, we first fit each model to two datasets with a compositional outcome and explanatory variable: the Education dataset (Section \ref{sec:educ}) and the White Cells dataset (Section \ref{sec:whitecells}). For the \citet{alenazi2019} model, we let $t(\bx) = ilr(\bx)$. These fitted coefficients are then used as the true coefficient values for each model when simulating data. Compositional covariates $\bx_{i}\ (i=1,\ldots, N; N=100, 250, 500, 1000)$ were simulated independently such that $x_{i} \sim Dirichlet(1, 1, 1)$. Because our direct regression model and the \citet{alenazi2019} model both directly specify $E[\by_{i} | \bx_{i}]$, we used the coefficients for each model from the two datasets to obtain the true conditional expected values, and then simulated $\by_{i} | \bx_{i} \sim Dirichlet(10 \cdot E[\by_{i} | \bx_{i}])$ for each model. For the \citet{chen2017} model, we simulated $ilr(\by_{i}) | \bx_{i} \sim \mathcal{N}(E[ilr(\by_{i}) | \bx_{i}], 1)$, and used $\by_{i} = ilr^{-1}(\by_{i})$ as the compositional outcome. 

Each of the three models were fit on the simulated data. To compare models, we generated a large, independent test set and obtained the true $E[\by_{i} | \bx_{i}]$ for each observation. We then obtain the average KLD between the true and estimated conditional means in this independent set. This full process is repeated 10,000 times for every combination of N, true data generating mechanism, and dataset. 

\begin{figure}[H]
\centering
\includegraphics[width=\linewidth]{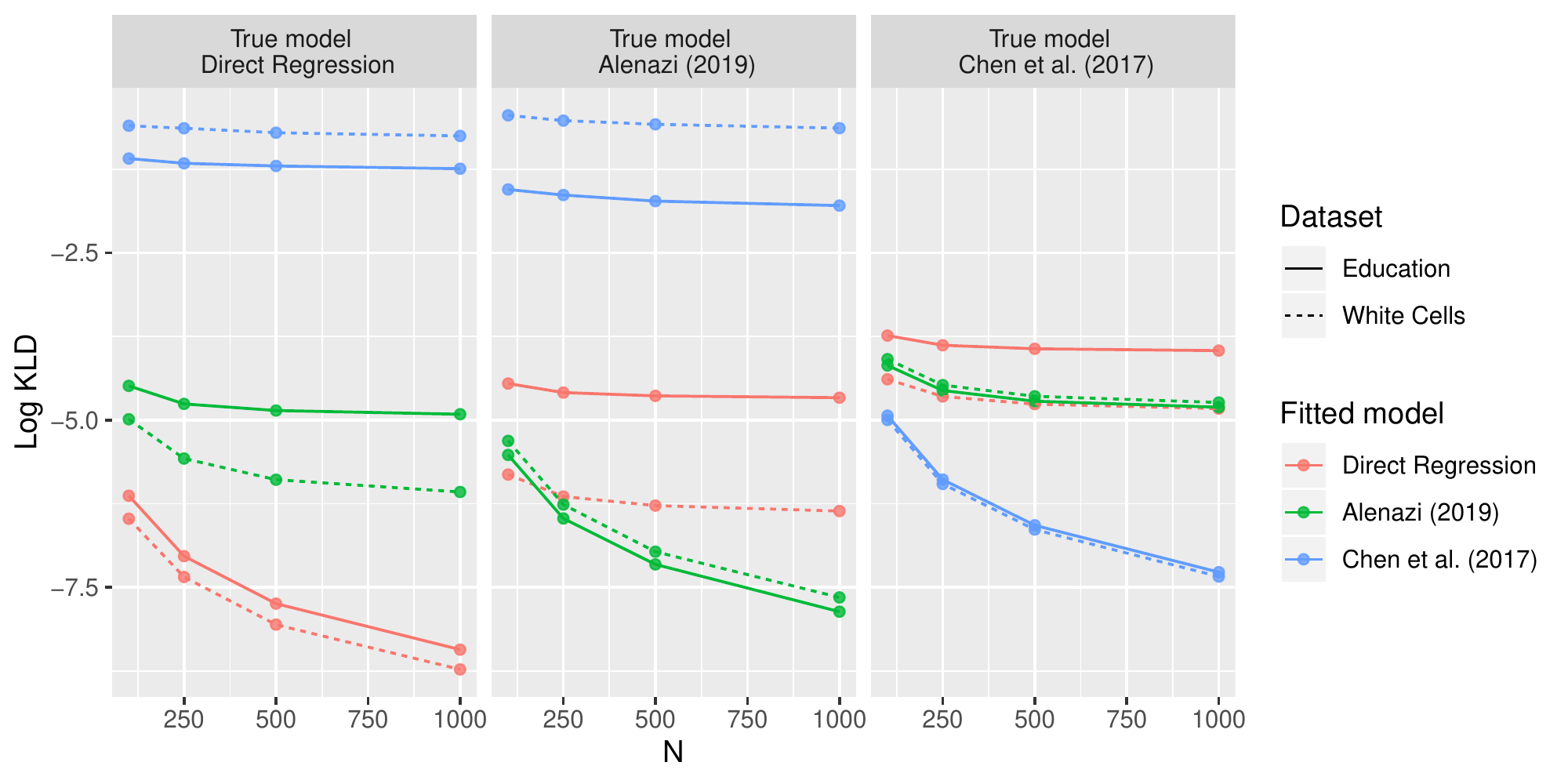}
\caption{Log KLD estimated using a test set, across various sample sizes and true models. Each column represents a different true model for the compositional outcome, with two sets of true coefficients values estimated on different datasets (solid and dashed lines). Each color shows the estimated Log KLD based on the fitted model.}\label{fig:comparemodels}
\end{figure}

For ease of comparison, Figure \ref{fig:comparemodels} shows the log KLD for each simulation setting, averaged across all 10,000 simulations. Unsurprisingly, the correctly specified model performs the best in conditional mean estimation across almost all settings.  Interestingly, the \citet{chen2017} model appears to perform much worse when it is misspecified, as compared to the direct regression model and the \citet{alenazi2019} model. Overall, these results show that each of these models can be used to model compositional regression models, and that the KLD (either estimated on a test set or through cross-validation) is a valid metric for model comparison. 

\subsection{Direct regression on different data generating mechanisms}\label{sec:diffdgms}

Because the direct regression model does not specify a likelihood for $\by | \bx$, we compare performance of the direct regression model across different data generating mechanisms that share the same conditional mean model. As in Section (\ref{sec:modelcompare}), we estimate the coefficients of the direct regression model on the same two datasets, and generate covariates $\bx_{i}$ using a uniform Dirichlet distribution.  We then generated $\by_{i} | \bx_{i}$ using three data generating mechanisms presented by \citet{murteira2016}:

\begin{enumerate}
    \item Dirichlet: The compositional outcome $\by_{i}$ is directly generated via the model $\by_{i} | \bx_{i} \sim Dirichlet(10 \cdot \bB^{'}\bx_{i})$.
    
    \item Multinomial (proportion): We first generate an individual ``sample-size'' $n_{i}$ from a  $Discrete-Uniform(1, 30)$ distribution. Individual counts are generated via $\by_{i}^{*} | \bx_{i} \sim Multinomial(n_{i},\bB^{'}\bx_{i})$, and the compositional outcome $\by_{i} $ is defined such that $y_{ik} = \frac{y_{ik}^{*}}{\sum_{k=1}^{3}y_{ik}^{*}}$.
    
    \item Dirichlet-multinomial (proportion): We introduce over-dispersion into the multinomial data generating scheme, by first simulating $\mathbf{p}_{i} | \bx_{i} \sim Dirichlet(10 \cdot \bB^{'}\bx_{i})$. Rather than simulating $\by_{i}^{*} | \bx_{i} \sim Multinomial(n_{i},\bB^{'}\bx_{i})$, we instead simulate $\by_{i}^{*} | \bx_{i} \sim Multinomial(n_{i},\mathbf{p}_{i})$. The compositional outcome $\by_{i} $ is again defined such that $y_{ik} = \frac{y_{ik}^{*}}{\sum_{k=1}^{3}y_{ik}^{*}}$.
\end{enumerate}

\begin{figure}[H]
\centering
\includegraphics[width=\linewidth]{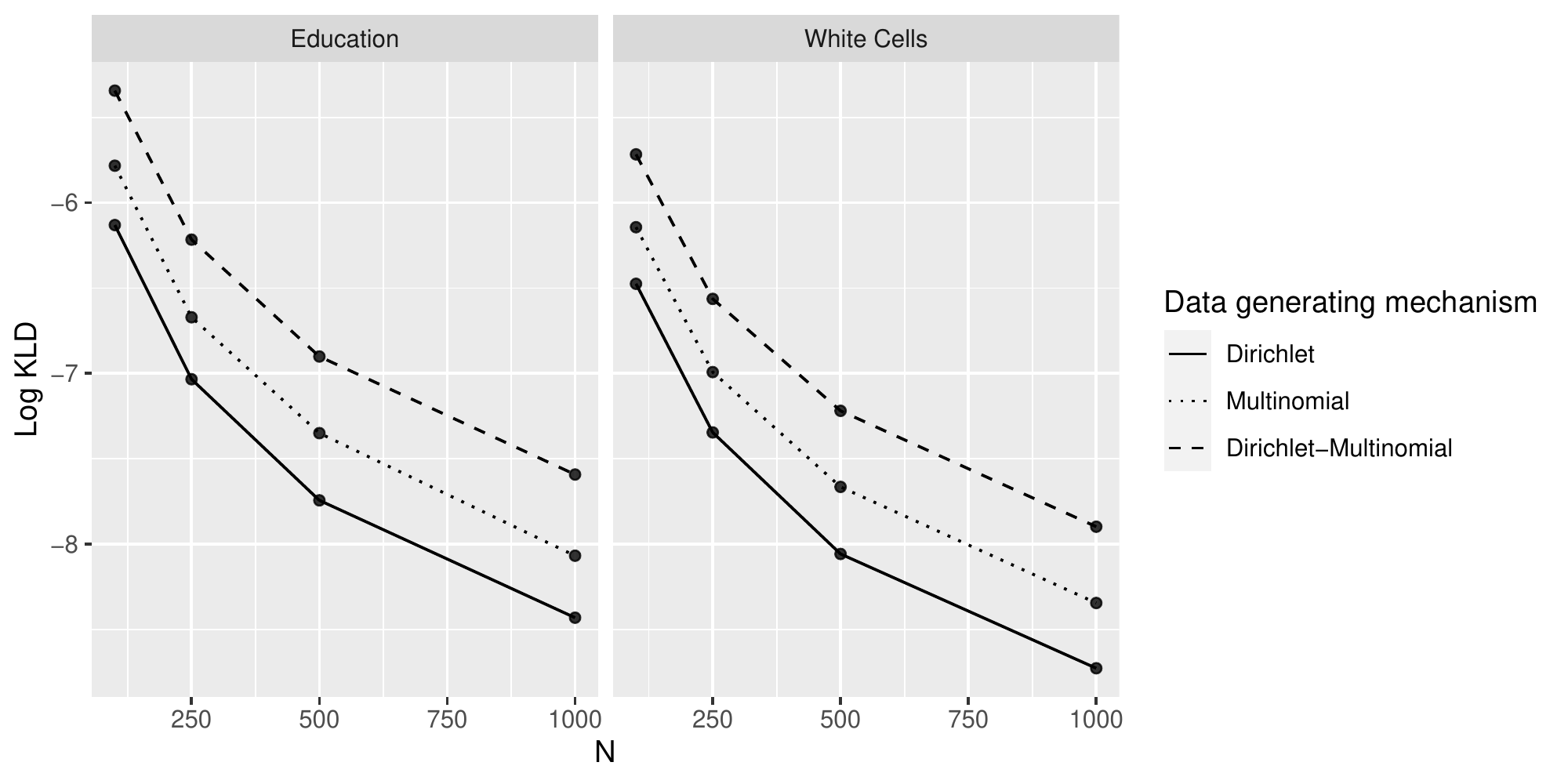}
\caption{KLD estimated using a test set, across various sample sizes and data generating mechanisms, with the conditional mean specified via the direct regression model. Each column represents a different true value for $\bB$, based on the two different real-world datasets. Each line type shows the estimated KLD for different data generating mechanisms for the compositional outcome.}\label{fig:diffdgms}
\end{figure}

The fitted direct regression models are evaluated via KLD on a test set, as in Section (\ref{sec:modelcompare}). Figure \ref{fig:diffdgms} shows that while the (log) KLD is similar across all data generating mechanisms, the model performs slightly worse for the models with higher variance for the compositional outcome. However, when the other two (incorrectly specified) models are fit to this simulated data, the direct regression model outperforms these models across all data generating mechanisms (Figure \ref{fig:appfig1}), again showing the importance of correctly specifying the conditional mean for the compositional outcome.

\subsection{Evaluating the Type-I and Type-II error rates of the global linear independence test}

To evaluate the testing procedure introduced in Section \ref{sec:indeptest} in terms of Type-I and Type-II error rates, we perform a simulation study that we detail in Web Appendix C. In summary, we observe that when $\by$ is linearly independent of $\bx$, our procedure produces well-calibrated Type-I error rates, regardless of the data generating mechanism for $\by$. We also observe that the permutation test generally has high power to detect linear relationships between E[$\by$] and $\bx$, except the case with small sample size (n=100) with weak linear relationship which is expected. Finally, we observe that when the true conditional mean is that specified by the direct regression model, but the model of \citet{chen2017} is fitted to the data, the permutation test unsurprisingly has lower power to detect dependence between $E[\by]$ and $\bx$.

\section{Applications}
To show that our method can realistically use data to address scientific questions in an interpretable manner, we now apply our method to two datasets which have a compositional predictor and a compositional outcome. 

\subsection{Educational status of mothers and fathers in European countries}\label{sec:educ}

Parental educational attainment has a large effect on child outcomes \citep{dubow2009long}. \citet{appliedCDA} provide a dataset that contains the percent of fathers and mothers with low, medium, and high education levels in 31 European countries. The question of interest presented by \citet{appliedCDA} is how the percentage of fathers with a given education level relate to the percentage of mothers with different education levels, across the 31 countries. We let $y_{ik}$ be the percentage of fathers with education level $k$ (1 = low (pre-primary, primary or lower secondary education), 2 = medium (upper secondary education and post-secondary non-tertiary education), 3 = high (first stage of tertiary education and second stage of tertiary education)) \citep{eurostat} in country $i$, and $x_{ij}$ be the percentage of mothers with education level $j$.

\begin{figure}[H]
\centering
\includegraphics[width=\linewidth]{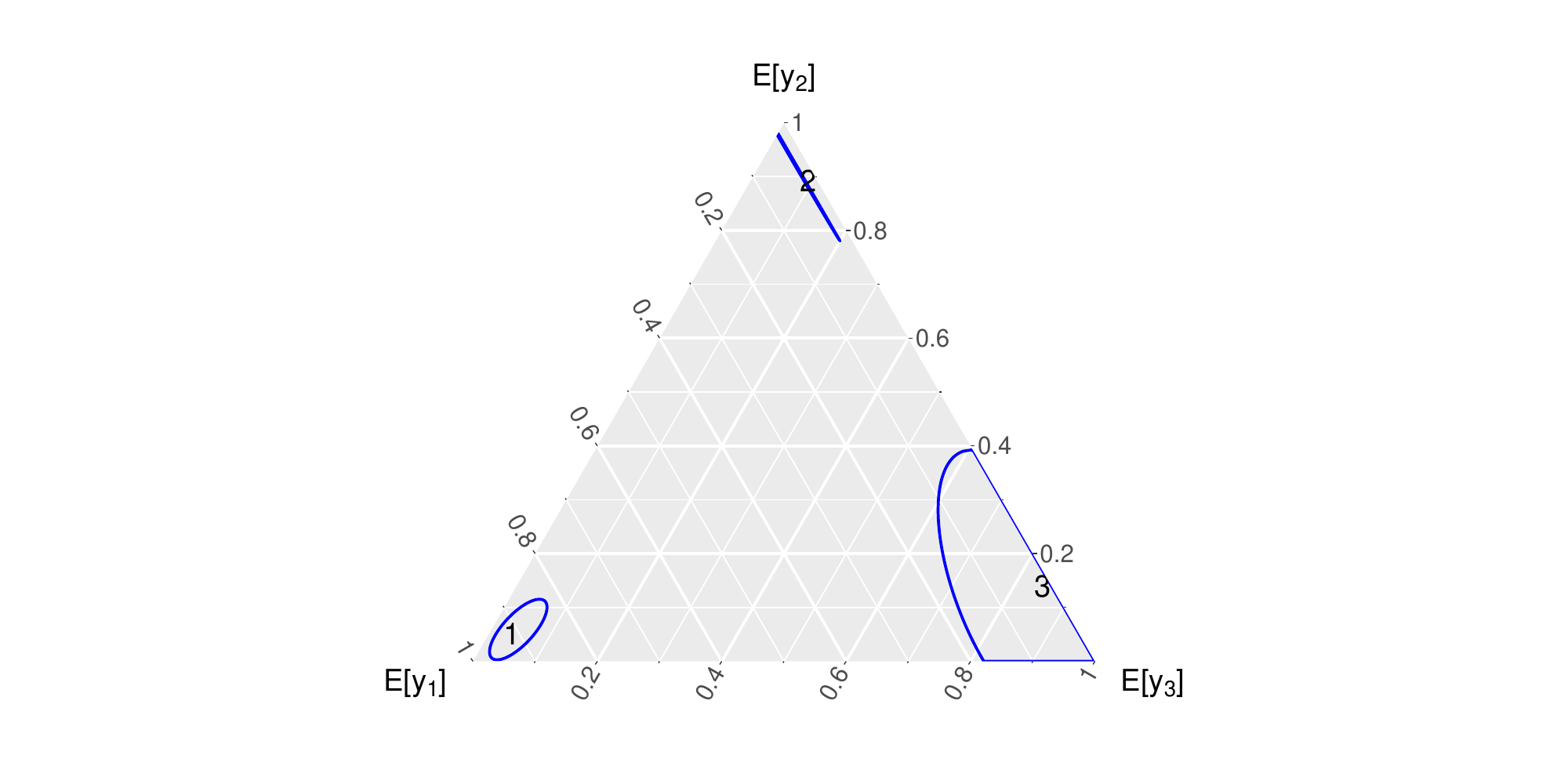}
\caption{Visualization of the coefficients for regression the percentage of fathers of a given education level on the percentage of mothers of a given education level. Each row of $\hat{\bB}$ is labeled with a number in the ternary diagram. The 95\% confidence region for each row is drawn in blue.}\label{fig:educationcoef}
\end{figure}

Fitting the model in (\ref{eq:markov}) leads to the following estimate of $\bB$:

\begin{equation*}
    \hat{\bB} = \begin{pmatrix}
    .91 & .05 & .04\\
    .00 & .91 & .09\\
    .00 & .14 & .86
    \end{pmatrix}
\end{equation*}
which shows high correlation between the educational attainment status of fathers and mothers (independence test p-value=0). The coefficients and 95\% confidence regions, obtained via bootstrap, are shown in Figure \ref{fig:educationcoef}. There is noticeably more uncertainty in estimation of $\bB_{3*}$ than in the other rows of $\bB$. In addition, there is very little uncertainty in $\hat{\bB}_{2,1}$.

The analytical interpretation of $\hat{\bB}$ means that increasing the percentage of mothers with a medium level of education level by .10, while decreasing the percentage of mothers with a low level of education level by .10, is associated with a change in the percentage of fathers with low, medium, and high educational status of -.091, .086, and .005, respectively. Similar affects are seen for other changes of the percentage of mothers with a given educational status. 

\begin{figure}[h!]
\centering
\includegraphics[width=\linewidth]{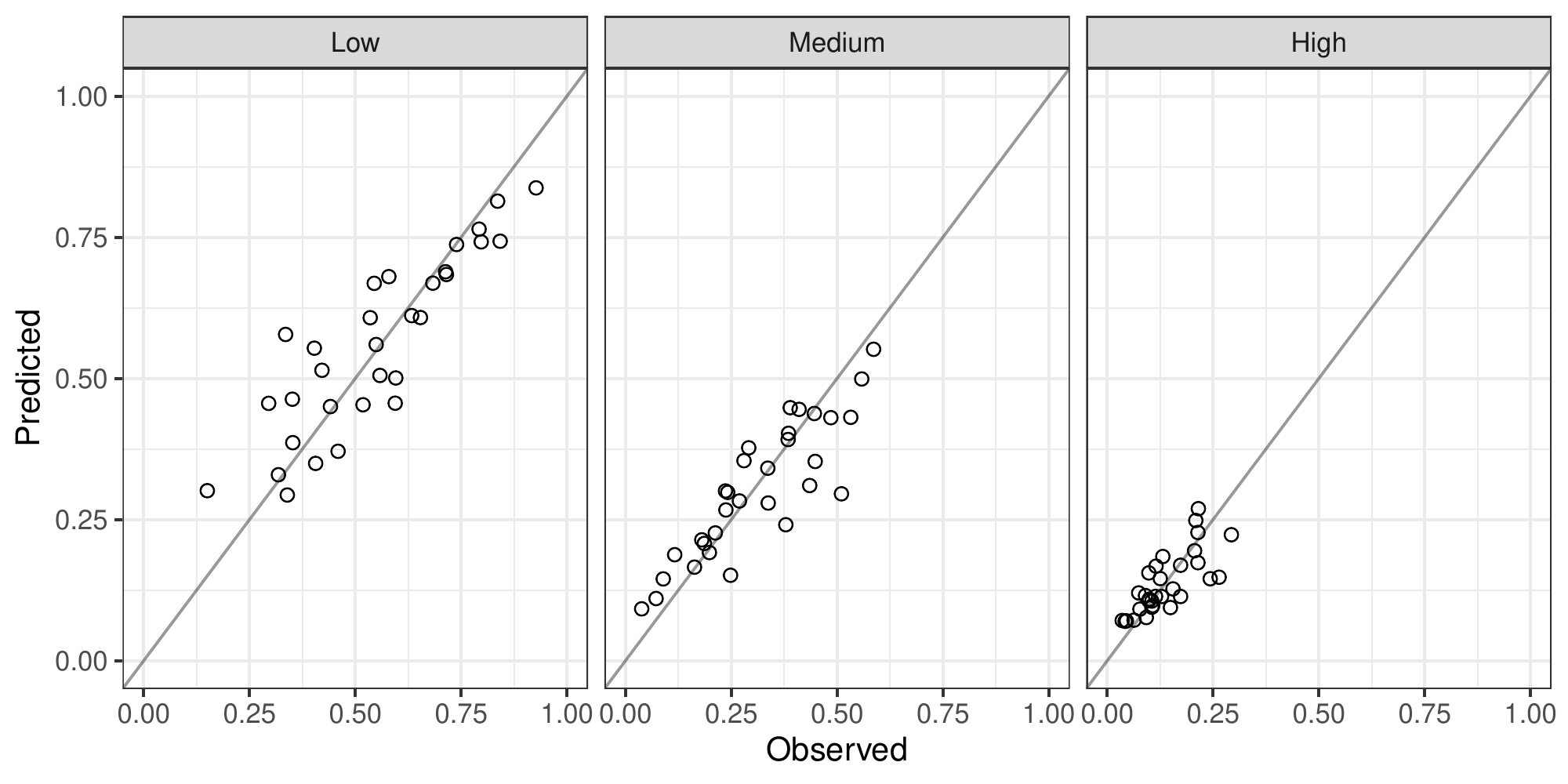}
\caption{Observed versus predicted father educational attainment compositions across each of the 31 countries based on leave-one-out analysis. The grey line represents the identity line.}\label{fig:education}
\end{figure}

To visualize the model fit, we first obtain predicted values for each of the father educational compositions, using leave-one-out cross-validation (LOOCV) \citep{friedman2001}, based off the mother educational compositions in each country. Figure \ref{fig:education} shows the observed versus predicted percentage of fathers with each level of education, across the 31 countries. The predicted percentages are all very close to the observed percentages, showing that our simple model is not only interpretable, but also appears to fit the observed data well.

We also compare our model to the models presented by \citet{chen2017} and \citet{alenazi2019} using the KLD between the observed $\by$ and predicted $\hat{\by}$, where $\hat{\by}$ is estimated via LOOCV for all three methods. Each of the three methods had a KLD of .024, indicating similar model fit. 

\subsection{White cell composition analysis}\label{sec:whitecells}

\citet{aitchison2003} and \citet{alenazi2019} consider a dataset in which the proportions of white blood cell types (granulocytes, lymphocytes, and monocytes) in 30 blood samples are determined by both a time-consuming microscopic analysis and an automated image analysis. The microscopic analysis is known to produce accurate results, while the accuracy of the image analysis is unknown. If the estimated compositions from the microscopic analysis can be predicted by the compositions estimated by the image analysis, it would be time-saving to use the automated image analysis in the future.

\begin{figure}[H]
\centering
\includegraphics[width=\linewidth]{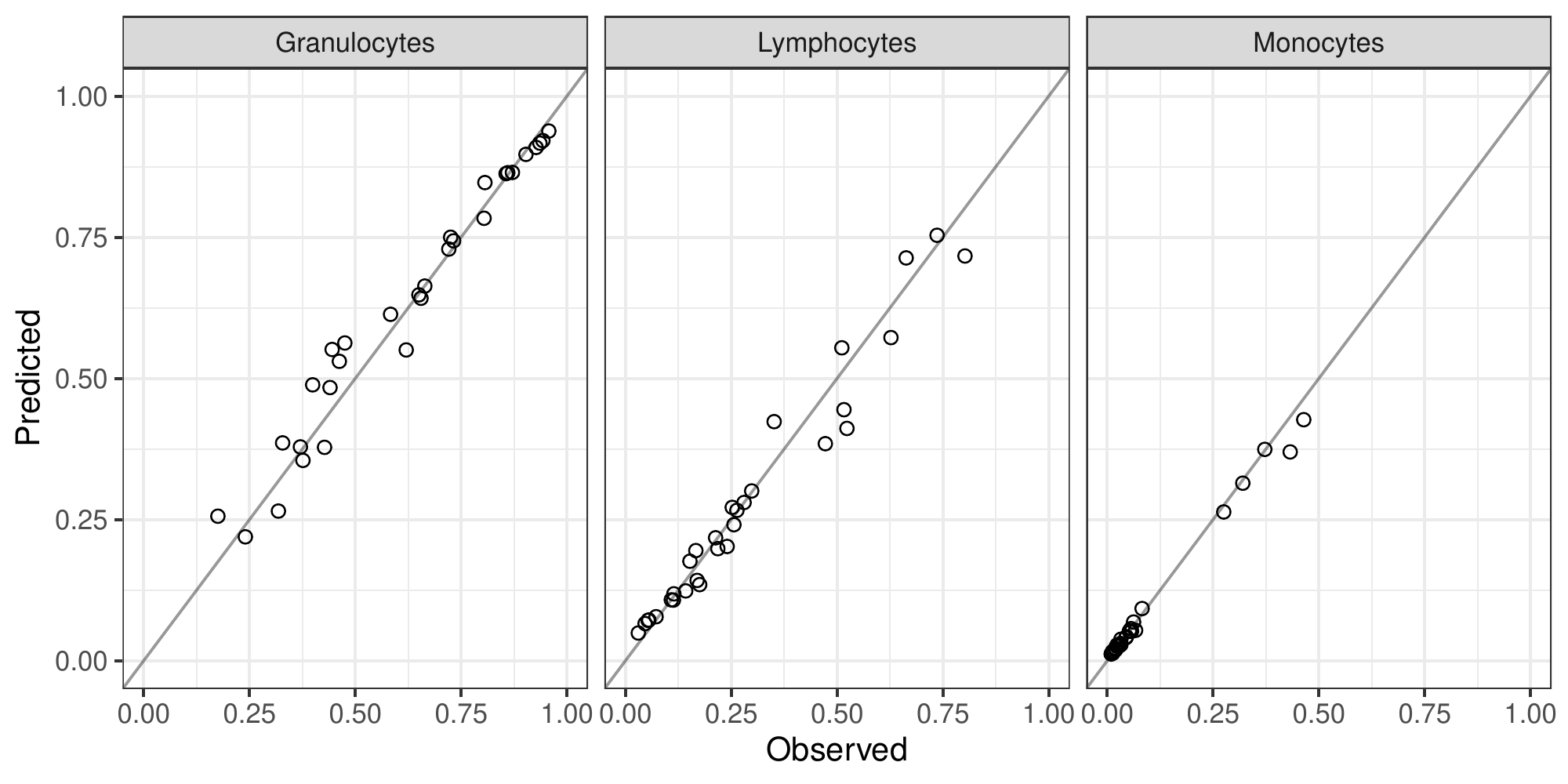}
\caption{Observed versus predicted white blood cell composition estimates using the microscopic analysis from each of the 30 samples using leave-one-out analysis. The grey line represents the identity line.}\label{fig:whitecells}
\end{figure}

We let $y_{ik}$ and $x_{ij}$ be the estimated composition of white blood cell type $k$ and $j$ (1 = granulocytes, 2 = lymphocytes, 3 = monocytes) by the microscopic and image analysis, respectively. The estimate of $\bB$ from our direct regression is

\begin{equation*}
    \hat{\bB} = \begin{pmatrix}
    .97 & .03 & .00\\
    .00 & 1.00 & .00\\
    .00 & .04 & .96
    \end{pmatrix}
\end{equation*}
which shows extremely high correlation between the compositional outcome and explanatory variables (independence test p-value=0). An increase in the estimated percentage of lymphocytes by .10 from the image analysis, at the expense of a .10 decrease of the estimated percentage of monocytes, is associated with a change in the estimated proportions of granulocytes, lymphocytes, and monocytes of 0, .096, and -.096, respectively, from the image analysis. Because $\hat{\bB}$ is extremely close to the identity matrix (i.e. perfect correlation), visualization of $\hat{\bB}$ provides little additional benefit in interpretation and we do not plot $\hat{\bB}$ in a ternary diagram. Figure \ref{fig:whitecells} again shows that our method produces extremely accurate predictions, obtained via LOOCV.

Finally, we again compare our method to the methods presented in Section (\ref{sec:review}) using the KLD. As in the analysis in (\ref{sec:educ}), the models perform nearly identically, with direct regression model and the model from \citet{chen2017} producing a KLD of .005, and the model from \citet{alenazi2019} producing a KLD of .006. These two analyses show that our method is not only more interpretable, but also comes without loss of fidelity to the observed data.

\section{Discussion}
In this manuscript, we have introduced a simple and novel direct regression model for compositional outcomes and explanatory variables that is fundamentally different from the existing suite of transformation-based methods for such problems. This direct regression model offers a simple interpretation of the regression coefficients, as opposed to the transformation-based methods. This simple interpretation will  facilitate the use of this model by practitioners who are not deeply familiar with complex compositional data transformations like the $ilr$, without having to resort to graphical techniques for visualizing the response surface. In addition to its simplicity, the direct regression model accommodates 0s and 1s in the data, seamlessly agress to aggregation of categories for both the covariate and the outcome, subsumes common structures like 2-way contingency tables, and and discrete time first order Markov processes. The estimating equations approach makes the model robust to misspecified data distributions. Fast parameter estimation is obtained through a likelihood-free EM algorithm, and a global null hypothesis test is proposed via a quasi-likelihood ratio test. Analysis of two datasets demonstrated how our model can accurately approximate observed scientific data generating mechanisms. 

One important future direction is developing a robust workflow for model comparison and selection for compositional regression problems. Although we have shown the potential of comparing the estimated KLD between models, there may be additional graphical and analytical tools that may yield better insight. Another important future direction is extending the direct regression model to allow for either continuous covariates or multiple compositional covariates, while maintaining simple interpretations for the compositional covariate coefficients. Current models for this problem simply extend the \citet{chen2017} model by including the continuous covariates in the model \cite{morais2018}. A potential solution is to use the direct regression model to model the partial dependence \citep{greenwell2017} between the compositional outcome and the compositional covariates of interest, but we leave this for future work.

\bibliography{references}  

\setcounter{table}{0}
\setcounter{figure}{0}
\renewcommand{\thefigure}{S\arabic{figure}}
\renewcommand{\thetable}{S\arabic{table}}

\section*{Supplementary Material}
\subsection*{Proof of Theorem 1}

We adopt this proof from the proof of Theorem 2.1 in \citet{yao2013}. For $i=1,\ldots, N$ and $k=1,\ldots, D_1$, let $z_{ik}^{(t+1)}$ be a discrete random variable such that

$$
P\left(z_{ik}^{(t+1)} = \frac{\tB_{jk}^{(t+1)}}{\tB_{jk}^{(t)}} \right) =  \frac{x_{ij}\tB_{jk}^{(t)}}{\sum_{j=1}^{D_{s}}x_{ij}\tB_{jk}^{(t)}} = \pi_{ijk}^{(t+1)}, j=1,\ldots, D_{s} \ .
$$

We then have

\begin{align*}
    f(\tB^{(t+1)}) - f(\tB^{(t)}) &= \sum_{i=1}^{N}\sum_{k=1}^{D_{r}}y_{ik}\log\left(\frac{\sum_{j=1}^{D_{s}}B_{jk}^{(t+1)}x_{ij}}{\sum_{j=1}^{D_{s}}B_{jk}^{(t)}x_{ij}}\right) \\
    &= \sum_{i=1}^{N}\sum_{k=1}^{D_{r}}y_{ik}\log\left(\sum_{j=1}^{D_{s}}\frac{B_{jk}^{(t)}x_{ij}}{\sum_{j=1}^{D_{s}}B_{jk}^{(t)}x_{ij}} \cdot \frac{B_{jk}^{(t+1)}x_{ij}}{B_{jk}^{(t)}x_{ij}}  \right) \\
    &= \sum_{i=1}^{N}\sum_{k=1}^{D_{r}}y_{ik}\log\left(\sum_{j=1}^{D_{s}}\pi_{ijk}^{(t+1)} \cdot \frac{B_{jk}^{(t+1)}x_{ij}}{B_{jk}^{(t)}x_{ij}}  \right)  \\
    &= \sum_{i=1}^{N}\sum_{k=1}^{D_{r}}y_{ik}\log\left(E[z_{ik}^{(t+1)}]\right)\\
    &\geq \sum_{i=1}^{N}\sum_{k=1}^{D_{r}}y_{ik}E[\log\left(z_{ik}^{(t+1)}\right)]\\
    &= \sum_{i=1}^{N}\sum_{k=1}^{D_{r}}y_{ik}\sum_{j=1}^{D_{s}}\pi_{ijk}^{(t+1)}log\left( \frac{\tB_{jk}^{(t+1)}}{\tB_{jk}^{(t)}} \right) \\
    &= \sum_{i=1}^{N}\sum_{k=1}^{D_{r}}\sum_{j=1}^{D_{s}}y_{ik}\pi_{ijk}^{(t+1)}\left[\log\left(\tB_{jk}^{(t+1)} \right) -  \log\left(\tB_{jk}^{(t)} \right)\right] \ .
\end{align*}

Because the M-step in (\ref{eq:mstepobj}) is the same regardless of whether $\by$ is categorical or compositional, this implies that 

$$
\sum_{i=1}^{N}\sum_{k=1}^{D_{r}}\sum_{j=1}^{D_{s}}y_{ik}\pi_{ijk}^{(t+1)}\log\left(\tB_{jk}^{(t+1)} \right) \geq  \sum_{i=1}^{N}\sum_{k=1}^{D_{r}}\sum_{j=1}^{D_{s}}y_{ik}\pi_{ijk}^{(t+1)}\log\left(\tB_{jk}^{(t)} \right)\ .
$$

We thus have shown that $f(\tB^{(t+1)}) - f(\tB^{(t)}) \geq 0$, with $f(\tB^{(t+1)}) - f(\tB^{(t)}) > 0$ if  $Q(\bB^{(t+1)}| \bB^{(t)}) > Q(\bB^{(t)}| \bB^{(t)})$.

\subsection*{Comparison of model performance for the simulations in Section 6.2}

\begin{figure}[H]
\centering
\includegraphics[width=\linewidth]{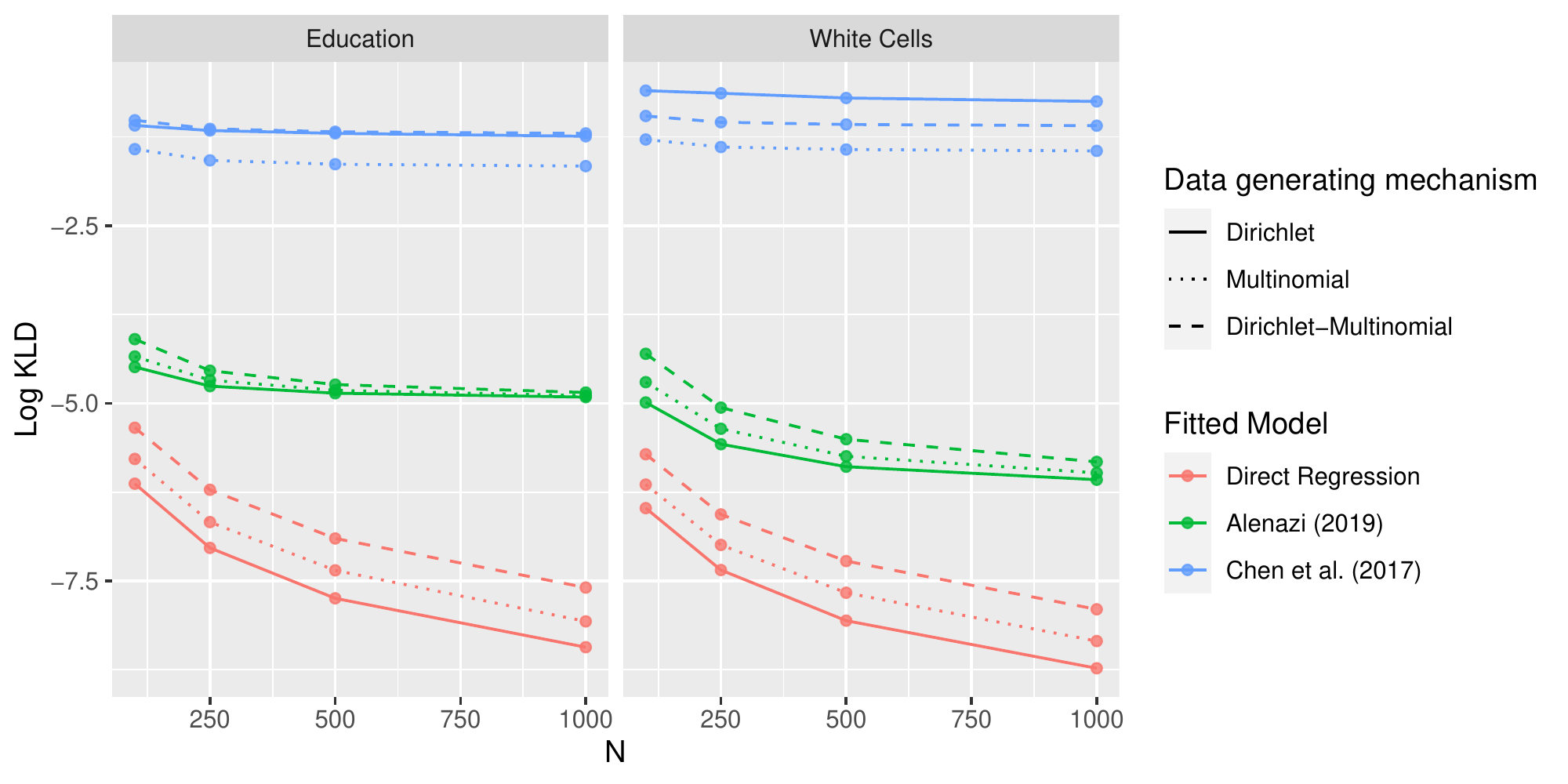}
\caption{Comparison of models via Log KLD, when the direct regression model specification is the correct conditional mean. The correctly specified direct regression model outperforms the other two models, across data generating mechanisms, coefficient values, and sample sizes.}\label{fig:appfig1}
\end{figure}

\subsection*{Simulation study to evaluate Type-I and Type-II error rates for the global independence test}

We again generated $\bx_{i}$ independently from a uniform Dirichlet distribution for $i=1,\ldots, N$, with $N=100, 250, 500, 1000$. We then generated $\by_{i} | \bx_{i}$ using the three data generating mechanisms introduced in Section \ref{sec:diffdgms}.

To evaluate the Type-I error rate, we generated data via the direct regression model by setting each row of $\bB$ to be $(\frac{1}{3},\frac{1}{3}, \frac{1}{3})$, which implies that $E[\by | \bx] = E[\by] = (\frac{1}{3},\frac{1}{3}, \frac{1}{3})$. We then simulated 10,000 data sets for each combination of of the 3 data generating mechanisms and 4 sample sizes. Table 2 \ref{tbl:type1} shows the percentage of the simulations where the observed p-value was below .05. Across all the sample sizes and data generating mechanisms for $\by$, we see that all observed Type-I error rates are very close to the nominal .05 rate, showing that the permutation test is well calibrated. 

\begin{table}
\caption{Empirical Type-I error rates across different sample sizes and data generating distributions for $\by$.}
\label{tbl:type1}
\centering
\begin{tabular}{|l|l|l|l|l|}
\hline
True Distribution     & N=100 & N=250 & N=500 & N=1000 \\ \hline
Dirichlet             & .050   & .052  & .051  & .052   \\ \hline
Multinomial           & .054   & .050  & .052  & .047   \\ \hline
Dirichlet-Multinomial & .050  & .050  & .048  & .052   \\ \hline
\end{tabular}
\end{table}

For evaluating the Type-II error rate when the direct regression model is correctly specified, we used the three different values for $\bB$:

\begin{equation*}
    \bB^{(1)} = \begin{pmatrix}
.90 & .05 & .05\\
.05 & .90 & .05\\
.05 & .05 & .90
\end{pmatrix};\ \bB^{(2)} = \begin{pmatrix}
.40 & .30 & .30\\
.30 & .40 & .30\\
.30 & .30 & .40
\end{pmatrix};\ \bB^{(3)}=\begin{pmatrix}
.90 & .05 & .05\\
.33 & .33 & .33\\
.33 & .33 & .33
\end{pmatrix}
\end{equation*}

The interpretations of $\bB^{(1)}$ and $\bB^{(2)}$ were introduced in Section \ref{sec:compreg}. $\bB^{(3)}$ represents the setting when $y_{1}$ and $x_{1}$ are highly correlated, but increasing $x_{2}$ at the expense of $x_{3}$ (and vice-versa) do not lead to any changes in $E[\by]$.

Table \ref{tbl:typeii} shows the percentage of simulations for each setting where the observed p-value was greater than .05. For $\bB^{(1)}$ and $\bB^{(3)}$, the permutation test shows extremely good performance in terms of Type-II error. Because the rows of $\bB^{(2)}$ are fairly close to being equal, the method unsurprisingly has a high Type-II error rate for $N=100$. Interestingly, the Type-II error rates differ across the three data generating mechanisms. As $N$ increases, the Type-II error rate decreases across all data generating mechanisms, with a Type-II error rate close to 0 when $N=1000$. 

\begin{table}[H]
\caption{Type-II error rates for the direct regression model across different values of $\bB$, data generating mechanisms, and sample sizes.}
\label{tbl:typeii}
\centering
\begin{tabular}{|l|l|l|l|l|l|}
\hline
Value for $\bB$            & True Distribution     & N=100 & N=250 & N=500 & N=1000 \\ \hline
\multirow{3}{*}{$\bB^{(1)}$} & Dirichlet             & .000   & .000     & .000   & .000    \\ \cline{2-6} 
                           & Multinomial           & .000   & .000     & .000   & .000    \\ \cline{2-6} 
                           & Dirichlet-Multinomial & .000   & .000   & .000   & .000    \\ \hline
\multirow{3}{*}{$\bB^{(2)}$} & Dirichlet             & .582   & .152   & .006   & .000    \\ \cline{2-6} 
                           & Multinomial           & .696   & .322   & .049   & .001    \\ \cline{2-6} 
                           & Dirichlet-Multinomial & .812   & .549   & .211   & .018    \\ \hline
\multirow{3}{*}{$\bB^{(3)}$} & Dirichlet             & .000   & .000   & .000   & .000    \\ \cline{2-6} 
                           & Multinomial           & .000   & .000   & .000   & .000    \\ \cline{2-6} 
                           & Dirichlet-Multinomial & .003   & .000   & .000   & .000    \\ \hline
\end{tabular}
\end{table}

We also evaluated the Type-II error rate of our method when the true model is the \citet{chen2017} model. We specify $E[ilr(\by_{i})_{k}]$ via the model in (\ref{eq:ilrreg}) using the following coefficient values:\\
\textbf{Model 1}
\begin{align*}
    &\beta_{01} = 1,\ \beta_{11} = 2,\ \beta_{21} = -1\\
    &\beta_{02} = -2,\ \beta_{12} = -1,\ \beta_{22} = 2
\end{align*}
\textbf{Model 2}
\begin{align*}
    &\beta_{01} = 1,\ \beta_{11} = .333,\ \beta_{21} = -.333\\
    &\beta_{02} = -2,\ \beta_{12} = -.333,\ \beta_{22} = .333
\end{align*}
\textbf{Model 3}
\begin{align*}
    &\beta_{01} = 1,\ \beta_{11} = 2,\ \beta_{21} = 0\\
    &\beta_{02} = -2,\ \beta_{12} = -1,\ \beta_{22} = 0
\end{align*}

Outcomes $\by_{i}$ were generated by first simulating $ilr(\by_{i})_{k} \sim \mathcal{N}(E[ilr(\by_{i})_{k} | \bx_{i}], 1)$ and then setting $\by_{i} = ilr^{-1}(ilr(\by_{i}))$. The permutation test achieved a Type-II error rate of 0 for all sample sizes and coefficient values, showing robustness to incorrect specification.

Finally, we evaluate the Type-II error rate of a likelihood ratio permutation test using the \citet{chen2017} model. We use a normal likelihood for the ILR transformed outcomes, and estimate the coefficients and standard errors via maximum likelihood, as in \citet{chen2017}. When the ILR model is correctly specified, using the coefficient values specified in the appendix, the Type-II error rate is 0 across all sample sizes. However, when the true conditional mean is that specified by the direct regression model, comparing Table \ref{tbl:typeiiILR} to Table \ref{tbl:typeii} shows the ILR regression model to have lower power than the direct regression model. 

\begin{table}[H]
\caption{Type-II error rates for the \citet{chen2017} model, using different values of $\bB$, data generating mechanisms, and sample sizes.}
\label{tbl:typeiiILR}
\centering
\begin{tabular}{|l|l|l|l|l|l|}
\hline
Value for $\bB$            & True Distribution     & N=100 & N=250 & N=500 & N=1000 \\ \hline
\multirow{3}{*}{$\bB^{(1)}$} & Dirichlet             & .000   & .000     & .000   & .000    \\ \cline{2-6} 
                           & Multinomial           & .000   & .000     & .000   & .000    \\ \cline{2-6} 
                           & Dirichlet-Multinomial & .000   & .000   & .000   & .000    \\ \hline
\multirow{3}{*}{$\bB^{(2)}$} & Dirichlet             & .642   & .225   & .017   & .000    \\ \cline{2-6} 
                           & Multinomial           & .914   & .854   & .743   & .515    \\ \cline{2-6} 
                           & Dirichlet-Multinomial & .909   & .834   & .692   & .405    \\ \hline
\multirow{3}{*}{$\bB^{(3)}$} & Dirichlet             & .000   & .000   & .000   & .000    \\ \cline{2-6} 
                           & Multinomial           & .320   & .014   & .000   & .000    \\ \cline{2-6} 
                           & Dirichlet-Multinomial & .231   & .004   & .000   & .000    \\ \hline
\end{tabular}
\end{table}


\end{document}